\documentclass[twocolumn,twocolappendix,tighten]{aastex63}

\received{03 March 2025}
\revised{23 July 2026}
\submitjournal{ApJ}

\pdfoutput=1 
\usepackage{amsmath,amstext}
\usepackage[T1]{fontenc}
\usepackage[figure,figure*]{hypcap}

\usepackage{graphicx}
\usepackage{graphics}
\usepackage{color}
\usepackage{lipsum}
\usepackage[caption=false]{subfig}
\definecolor{dartmouthgreen}{rgb}{0.05, 0.5, 0.06}
\usepackage{afterpage}
\usepackage{natbib}
\usepackage[percent]{overpic}

\def\deg{\hbox{$^\circ$}}
\def\hr{\textsuperscript{h}}
\def\min{\textsuperscript{m}}
\def\sec{\textsuperscript{s}\hspace{-0.7mm}}

\def\chandra{\textit{Chandra}}
\def\xmm{\textit{XMM-Newton}}
\def\wise{\textit{WISE}}

\usepackage{rotating}

\graphicspath{ {Images/} }

\shorttitle{A New X-ray Transient in NGC 4945}
\shortauthors{Pfeifle et al.}
\begin{document}
\title{XMM-Newton and Swift Unveil Another X-Ray Transient in NGC 4945, XMM J130514.64-493311.27}


\author[0000-0001-8640-8522]{Ryan W. Pfeifle}
\altaffiliation{NASA Postdoctoral Program Fellow}
\affiliation{Oak Ridge Associated Universities, NASA NPP Program, Oak Ridge, TN 37831, USA}
\affiliation{X-ray Astrophysics Laboratory, NASA Goddard Space Flight Center, Code 662, Greenbelt, MD 20771, USA}

\author[0009-0008-4232-486X]{Kimberly A. Weaver}
\affiliation{X-ray Astrophysics Laboratory, NASA Goddard Space Flight Center, Code 662, Greenbelt, MD 20771, USA}

\author{Jenna M. Cann}
\altaffiliation{NASA Postdoctoral Program Fellow}
\affiliation{X-ray Astrophysics Laboratory, NASA Goddard Space Flight Center, Code 662, Greenbelt, MD 20771, USA}
\affiliation{Oak Ridge Associated Universities, NASA NPP Program, Oak Ridge, TN 37831, USA}

\author{Murray Brightman}
\affiliation{Cahill Center for Astrophysics, California Institute of Technology, 1216 East California Boulevard, Pasadena, CA 91125, USA}

\author{Miranda McCarthy}
\affiliation{Southeastern Universities Research Association, Washington DC 20005, USA}
\affiliation{Center for Research and Exploration in Space Science and Technology, NASA/GSFC, Greenbelt, MD 20771, USA}
\affiliation{X-ray Astrophysics Laboratory, NASA Goddard Space Flight Center, Code 662, Greenbelt, MD 20771, USA}





%


\begin{abstract}
NGC 4945 hosts a Compton-thick, low luminosity AGN and a nuclear starburst in its core and has been the subject of many previous X-ray investigations of the AGN emission structure, nuclear starburst properties, and ultraluminous X-ray (ULX) source populations. Indeed, NGC 4945 has proven to be a rich system for ULX and X-ray transients, with at least four X-ray transients being reported over the last 15 years. Here we report the detection of an X-ray transient source, XMM J130514.64-493311.27, in NGC 4945 in the latest and deepest XMM-Newton observation of NGC 4945 to-date. The source lies $\sim5.5'$ south of the nucleus and was not detected in any previous XMM-Newton or Chandra imaging, but the source was detected by Swift-XRT in 2008, 2019, and in 2022, the latter of which coincides roughly with our new XMM-Newton imaging; we are therefore tracing the most recent outburst of this object with XMM-Newton. The source is soft with $\Gamma\approx3$, and our spectroscopic analysis suggests the source is best characterized by a multicolor disk + power law model or, alternatively, a multicolor disk + thermal plasma model, with a 0.3-10\,keV X-ray luminosity of $\sim2.2-2.3\times10^{38}$. We find no evidence for a counterpart in the optical, near-infrared, ultraviolet, or in the radio, but we identify a candidate counterpart in NEOWISE mid-IR light curves. We propose that this X-ray source is an X-ray binary within NGC 4945 caught during its most recent active phase.
\end{abstract}


\section{Introduction} 
\label{sec:intro}
NGC 4945 is a unique astrophysical laboratory hosting not only a Compton-thick AGN -- exhibiting both parsec- and kiloparsec-scale iron K$\alpha$ nebulae \citep{2003ApJ...588..763D, 2017MNRAS.470.4039M,weaver2024} -- but also kiloparsec-scale AGN and starburst driven winds \citep[e.g.,][]{2002MNRAS.335..241S}. This nearby (D\,$\sim4$\,Mpc), edge-on galaxy hosts a rich set of star forming regions \citep{2020ApJ...903...50E} and X-ray point sources \citep{colbert2004,kaaret2008,liu2011,vulic2018}. Numerous studies over the last several decades have studied the nuclear and off nuclear X-ray source populations in NGC 4945, demonstrating that these X-ray sources are driven by ultraluminous X-ray sources \citep[ULXs,][]{brandt1996} and X-ray binaries \citep[XRBs,][]{kaaret2008,vulic2018}; several of these XRBs have since been further classified as black hole or neutron star X-ray binaries based on hard X-ray NuSTAR diagnostics, with $>70\%$ of the $4-25$ or $12-25$\,keV point source emission originating from neutron stars \citep[][]{vulic2018}. Prior work has also demonstrated the propensity for nuclear and off nuclear X-ray transients driven by ULXs \citep[e.g.,][]{ide2020,isobe2008,brightman2023} and supernovae \citep{monard2011,chakraborti2013}. In particular, \citet{brightman2023} concluded that NGC 4945 should exhibit transient ULXs at a rate of $0.7\pm0.5$ yr$^{-1}$, and their results suggested that transient ULXs may be the dominant population of ULXs within NGC 4945.  


We are conducting an in depth analysis of new XMM-Newton imaging of NGC 4945, obtained in order to provide exquisite data on the AGN outflow \citep[see, e.g.,][]{weaver2024}. Upon a preliminary examination of the 2022 XMM-Newton observation of NGC 4945, we noticed a bright, off nuclear source in NGC 4945 which was not seen in either of the 2001 or 2004 observations, nor has it been reported in any Chandra observation over the last 20 years. 

\begin{figure}
    \includegraphics[width=0.99\linewidth]{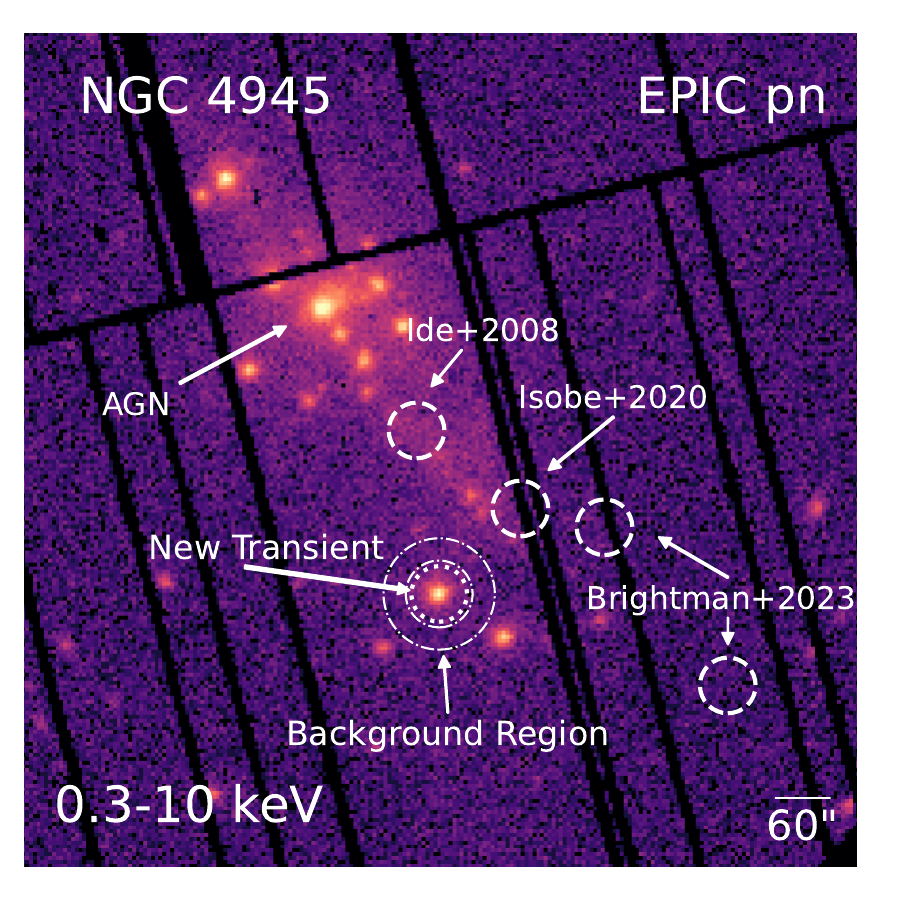}
    \caption{The XMM-Newton 0.3-10\,keV pn X-ray image of NGC 4945 from 2022. On the XMM-Newton image, we have overlaid a series of white dashed apertures to denote the positions of previously reported ultraluminous X-ray sources and X-ray binaries \citep[e.g.,][]{isobe2008,ide2020,brightman2023} and the position of the newly detected source studied in this work. The dotted 30'' aperture and dash-dotted 36''-60'' annulus represent the source and background extraction regions, respectively.}
    \label{fig:allsources}
\end{figure}

In this work we describe this X-ray transient, referred to as XMM J130514.64-493311.27, discovered in new XMM-Newton EPIC imaging of NGC 4945. The outline of this paper is as follows: in Section~\ref{sec:obs} we describe our data processing steps. Section~\ref{sec:analysis} details our XMM-Newton X-ray imaging and spectroscopic analysis of the new transient source, and discuss the activation of the source based on Swift-XRT imaging in Section~\ref{sec:swift} and the search for a multiwavelength counterpart  in Section~\ref{sec:multiwavelengthsource}. We summarize our results describe the nature of this new source in Section~\ref{sec:summ_discuss}. Throughout this work we adopt a luminosity distance of 3.6 Mpc to NGC 4945 \citep{karachentsev2002}; at this distance, 1'' corresponds to $\sim18$\,pc.

\section{Observations and Data Reduction}
\label{sec:obs}

NGC 4945 was targeted in three separate XMM-Newton observation programs over the last three decades: a 24\,ks observation on 21 Jan. 2001 (0112310301), a 65\,ks observation on 10 Jan. 2004 (ObsID 0204870101), and a $\sim180$\,ks observation split across four separate exposures, ObsIDs 0903540101 (5 July 2022), 0903540201 (16 August 2022), 0903540301 (5 July 2022), and 0903540401 (16 August 2022). ObsID 0903540301 and 0903540401 consist of CalClosed observations and contain no science events, while ObsID 0903540201 suffers from significant particle background flaring and does not exhibit a stable background. We therefore focus only on ObsIDs 0112310301 (24\,ks), 0204870101 (65\,ks), and 0903540101 (125\,ks) in this work. 

Each observation was reprocessed using SAS v20.0 \citep{gabriel2004} and Heasoft v6.30.1 \citep{blackburn1995}.\footnote{https://heasarc.gsfc.nasa.gov/ftools} Initial reprocessing was carried out using the \textsc{epproc} and \textsc{emproc} commands while the \textsc{evselect} command was used for pattern, bad pixel, and energy filtering. Background light curves were extracted using the \textsc{heasoft} \textsc{lcurve} command from a source-free event file with energies $10-12$\,keV for pn and $>10$\,keV for the mos cameras, and a good time interval file for each camera was generated using the \textsc{tabgtigen} command. Final background flare removal was performed using \textsc{evselect}. We generated science images binned to the native pixel size of pn (4.1'') and the MOS (1.1'') cameras for the 0.3-10\,keV, 0.3-2\,keV, and 2-10\,keV energy bands using \textsc{evselect}. A new transient source, XMM J130514.64-493311.27, was identified in the latest XMM-Newton observation; despite having sufficient depth to detect this source at its observed luminosity ($L_{2-10\,\rm{keV}}\approx7.1-7.8\times10^{37}$\,erg\,s$^{-1}$, see Section~\ref{sec:spectralfitting}), no prior XMM-Newton or Chandra observations of NGC 4945 detected this X-ray source (including stacked Chandra imaging, as discussed below in Section~\ref{sec:analysis}.) 

Using the \textsc{evselect} command, we extracted the spectrum of XMM J130514.64-493311.27 from the cleaned pn, mos1, and mos2 event files using a 30'' radius aperture centered on the source position (discussed below in Section~\ref{sec:analysis}). A background spectrum was drawn from an annulus centered on the source position with inner and outer radii of 36'' and 60'', respectively (see dash-dotted annulus in Figure~\ref{fig:allsources}). The source is situated at the edge of XMM-Newton's Cu hole \citep[see Figure 1 in][ for an illustration of the Cu hole; see also discussion in Section~\ref{sec:analysis}in this work]{weaver2024}, so this annulus acts as a compromise to sample the local instrumental, Galactic, and extragalactic X-ray backgrounds; a single background aperture placed further south (north) would undersample (oversample) the local background from the halo emission of NGC 4945 and would oversample (undersample) XMM-Newton's instrumental emission. Spectral responses were generated using the \textsc{rmfgen} and \textsc{arfgen} commands, and the spectrum from each camera was binned to 1 count per bin using the \textsc{heasoft} \textsc{grppha} command in order to fit the spectra in \textsc{xspec} \citep{arnaud1996} using Cash statistics \citep{cash1979}.

\section{Data Analysis}
\label{sec:analysis}

\subsection{XMM-Newton Imaging Analysis}
\begin{figure*}
    \includegraphics[width=1.0\linewidth]{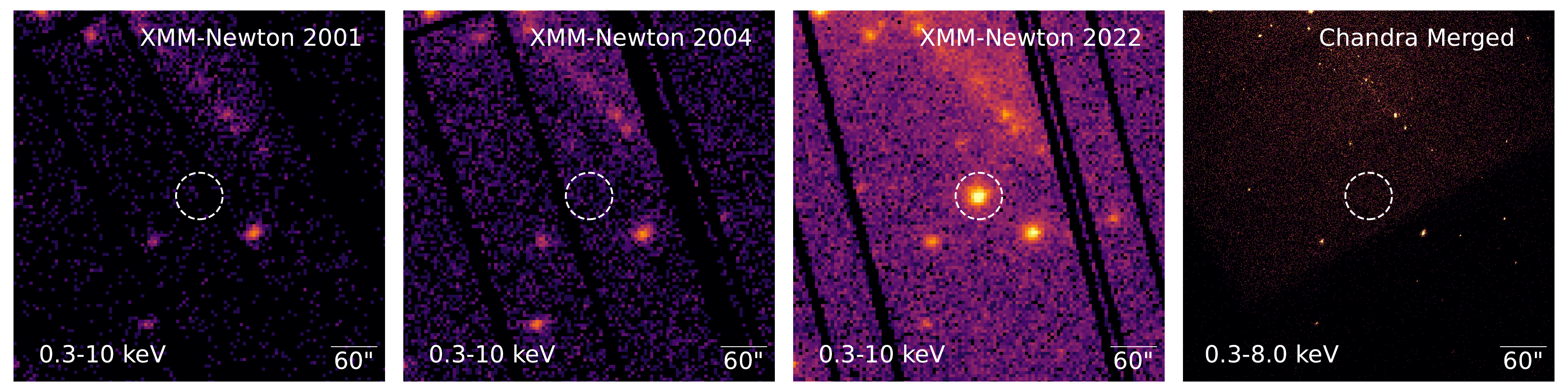}
    \caption{This four panel figure shows (from left to right) the XMM-Newton pn imaging from 2001, 2004, and 2022 along with the stacked mosaic of all available Chandra observations from 2000-2021. These panels are zoomed-in to show the position of the newly-detected off-nuclear X-ray source (dashed white circle). In each panel, the scale bar in the bottom right corner indicates 60'' while the energy band is indicated in the bottom left corner. These X-ray images are shown in log scale using the perceptually uniform color map `magma.'}
    \label{fig:xmmcompareepochs}
\end{figure*}

We identified a new off-nuclear X-ray transient, XMM J130514.64-493311.27, in NGC 4945 during our preliminary comparison of the XMM-Newton imaging from ObsID 0903540101 and the two previous images taken in 2001 and 2004 (ObsIDs 0112310301 and 0204870101). While the off-nuclear source is prominent in our most recent imaging, the source is not seen in either of the previous XMM-Newton images (Figure~\ref{fig:xmmcompareepochs}). We used the SAS command \textsc{edetectchain} to determine the precise coordinates of this new source within ObsID 0903540101, which we found to be: $\alpha=13\hr{}05'14.\hspace{-1mm}''64$, $\delta=-49\deg{}33'11.\hspace{-1mm}''27$, with a statistical uncertainty of 0.125'' and an astrometric uncertainty of $\sim1.3''$, the latter of which was determined via a cross-match between sources detected within the XMM-Newton FOV and the Gaia DR3 source catalog \citep{gaia2016,gaia2023}. We also ran \textsc{edetectchain} on the two previous XMM-Newton observations; \textsc{edetectchain} did not detect a source at this position in either of the prior observations. Clearly, this source turned on some time between 2004 and 2022. 

To investigate when XMM J130514.64-493311.27 may have become active, we retrieved all archival Chandra observations of NGC 4945 (from 2000 to 2021) and reprocessed them using the \textsc{chandra\_repro} command in CIAO \citep{fruscione2006}. We inspected each event file as well as the stacked mosaic event file visually and by running \textsc{wavdetect} (with wavelet scales of 2 4 8 16) to identify any putative sources in the individual event files and the stacked mosaic (shown in the fourth panel of Figure~\ref{fig:xmmcompareepochs}). In neither case did we find a \textit{Chandra} counterpart at the location of XMM J130514.64-493311.27. The position of this object unfortunately fell just outside of the field of view of the most recent 2021 Chandra observation (ObsID 24986), but the Chandra observation from 2018 (ObsID 21082) does cover this position. The Chandra Point Source Catalog 2.0\footnote{http://cda.cfa.harvard.edu/cscweb/index.do} \citep{evans2024} yields an upper limit of $F_{0.5-7\,\rm{keV}}<1.37\times10^{-15}$ erg\,cm$^{-2}$\,s$^{-1}$ (or $L_{0.5-7\,\rm{keV}}<2.1\times10^{36}$ erg\,s$^{-1}$) at the position of XMM J130514.64-493311.27. Given the lack of detection in the 2018 Chandra observation (in addition to the stacked event file), the source likely became active sometime between December 2018 and July 2022 (see our Swift-XRT analysis in Section~\ref{sec:swift}). 

\begin{figure}
    \centering
    \includegraphics[width=1.0\linewidth]{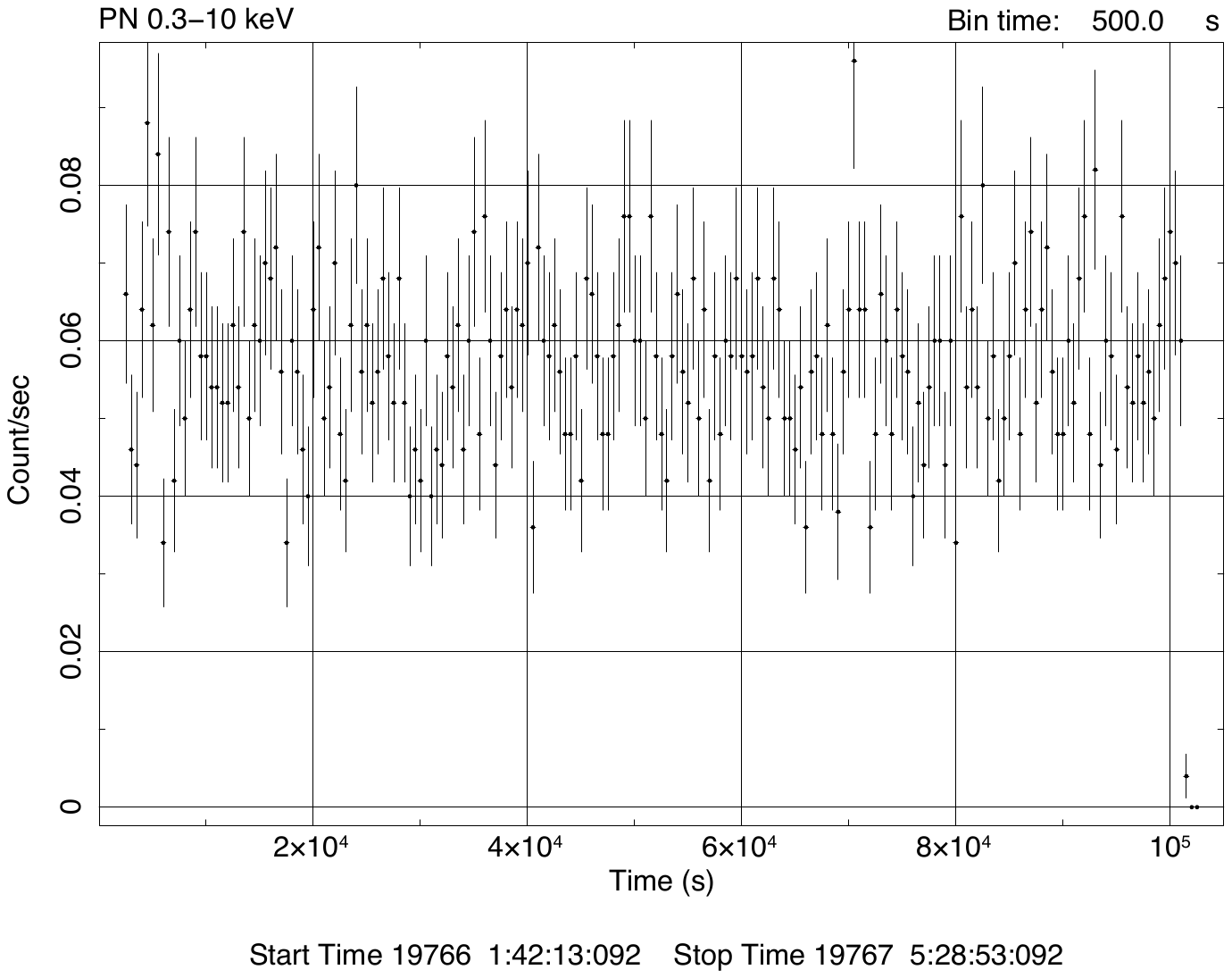}
    \caption{$\sim99$\,ks 0.3-10\,keV light curve of the new X-ray transient in NGC 4945. The source shows no obvious sign of flaring across the exposure time.}
    \label{fig:lightcurve}
\end{figure}

We further investigated whether XMM J130514.64-493311.27 was variable over the course of our observation. We extracted a 0.3-10\,keV light curve for the source from the pn imaging using the \textsc{lcurve} command in \textsc{heasoft} and binned the light curve at 100\,s, 500\,s, and 1000\,s intervals. From these binned light curves, we derived a fractional variability of 7.4\%, 12.6\%, and 13.8\%, respectively, consistent with no substantial variability or flaring across the observation. Figure~\ref{fig:lightcurve} displays the source light curve binned at 500\,s. We used the SAS command \textsc{eregionanalyse} to extract background-subtracted source counts corrected for the enclosed energy fraction of our extraction aperture. We detected $5647\pm93$, $4882\pm81$, and $774\pm45$ (background-subtracted) counts in the 0.3-10\,keV, 0.3-2\,keV, and 2-10\,keV bands. XMM J130514.64-493311.27 is soft; \textsc{eregionanalyse} computed a hardness ratio (HR) of -0.73 using the equation HR = $(H-S)/(H+S)$, where $H$ and $S$ are the counts in the 2-10\,keV and 0.3-2\,keV bands, respectively.

\subsection{X-ray Spectroscopy with XMM-Newton}
\label{sec:spectralfitting}

Due to the off-axis position of XMM J130514.64-493311.27 on the XMM-Newton detectors, the spectrum was contaminated by strong instrumental lines, particularly the Cu line at 8.1\,keV;\footnote{The Cu background is variable across the detector; the observed Cu emission in the source region may only approximately match the observed Cu emission in the background region.} since the spectrum of the source was already dominant in the soft X-rays, we elected to limit our spectroscopic analysis to the 0.3-7\,keV band. During the fitting process, we fit the spectra from the three \textsc{epic} cameras simultaneously. Each choice of spectral model included a constant component to account for inter-detector sensitivity that was left free to vary within $\pm30$\% of the first data instance loaded into \textsc{xspec}; these constants never exceeded $\sim10\%$.

Given the resemblance of XMM J130514.64-493311.27 to that of a black body, we began by fitting an absorbed black body model,  \textsc{tbabs}$\times$\textsc{diskbb}, where we left all of the parameters free to vary. We also tried two other models: an absorbed power law model (\textsc{tbabs}$\times$\textsc{pl}), and an absorbed thermal plasma model (\textsc{tbabs}$\times $\textsc{apec}), with all parameters left free to vary with the exception of the elemental abundance and redshift parameters in the thermal plasma model, which we assumed to be solar and $z=0.001797$, respectively. For any cases where the best fit model returned a column density lower than the known Galactic value ($2.2\times10^{21}$\,cm$^{-2}$) we instead froze the column density to the known Galactic value and began the fitting process again. The \textsc{apec} model alone could not reproduce the observed spectrum. The power law model (with steep photon index $\Gamma=2.97_{-0.08}^{+0.09}$) and multicolor disk model (with inner accretion temperature $0.56_{-0.01}^{+0.02}$\,keV and column density frozen to the Galactic value) both provided a reasonable fit between $\sim0.5-3$\,keV but showed deviations from the observed spectrum below 0.5\,keV and above 3\,keV. In the case of the multicolor disk model, the apparent inner accretion disk radius can be computed using the normalization via the expression $norm = (r_{\rm{in}}/D_{10})^2{\,\rm{cos}}(i)$, where D10 is the distance to the source in units of 10\,kpc and $i$ is the inclination angle of the disk. The true inner disk radius $R_{in}$ can be derived using the apparent disk radius via the expression $R_{\rm{in}}=\kappa^2\xi r_{\rm{in}}$, where $\xi=0.412$ is a correction factor for the inner boundary condition \citep{kubota1998} and $\kappa=1.7$ is a spectral hardening factor \citep{shimura1995}. Using these expressions, the simple multicolor disk model yielded an inner accretion disk radius of $R_{in}\,$cos$^{1/2}(i)=144.9$\,km. 

To try to account for emission observed below 0.5\,keV and above 3\,keV, we then tried four additional models that used the absorbed multicolor disk  model as a foundation: 
\begin{itemize}
    \item \textsc{tbabs}$\times$(\textsc{diskbb}+\textsc{bbody}), where both thermal temperatures were free to vary within the range 0.01-5\,keV.
    \item \textsc{tbabs}$\times$(\textsc{diskbb}+\textsc{apec}), referred to as the multicolor disk + thermal plasma model, where the \textsc{diskbb} and \textsc{apec} temperatures were free to vary within the ranges 0.01-5\,keV and 0.01-2\,keV.
    \item \textsc{tbabs}$\times$(\textsc{diskbb}+\textsc{pl}), referred to as the multicolor disk + power law model, where the thermal temperature and photon index were left free to vary within the ranges of 0.01-5\,keV and $\Gamma$=0.1-5, respectively.
\end{itemize}
The \textsc{diskbb+bbody} model can be physically interpreted as modeling the emission from the accretion disk via the multicolor disk component as well as emission from the surface of a neutron star via the blackbody component \cite[][]{morii2010,gambino2019}. \textcolor{black}{The \textsc{diskbb+pl} model can be interpreted as modeling emission from the accretion disk (via the multicolor disk component) as well as Compton up-scattered emission from a corona (via the power law component) \citep{done2007}.} The \textsc{diskbb+apec} model can be interpreted as modeling the emission from the accretion disk as well as soft X-ray emission potentially due to the nearby halo of NGC 4945 as well as Galactic and cosmic soft X-ray background emission.

\begin{figure}
    \centering
    \subfloat{\includegraphics[width=0.8\linewidth]{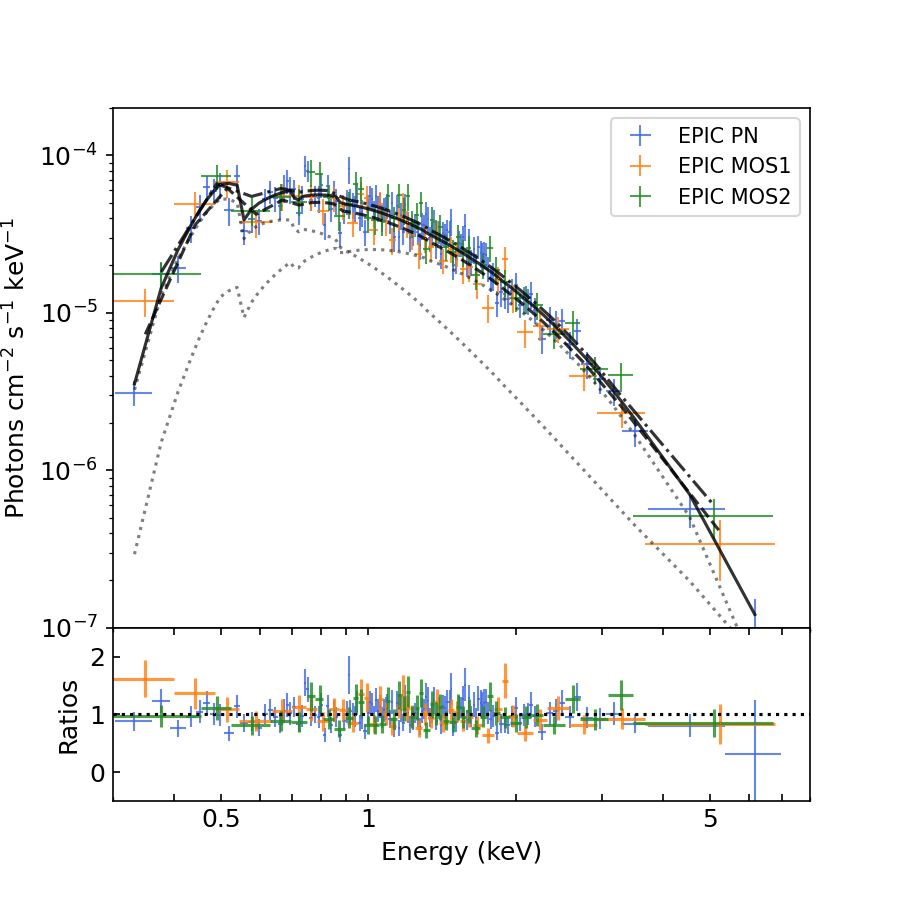}} \\ \vspace{-0.5cm}
    \subfloat{\includegraphics[width=0.8\linewidth]{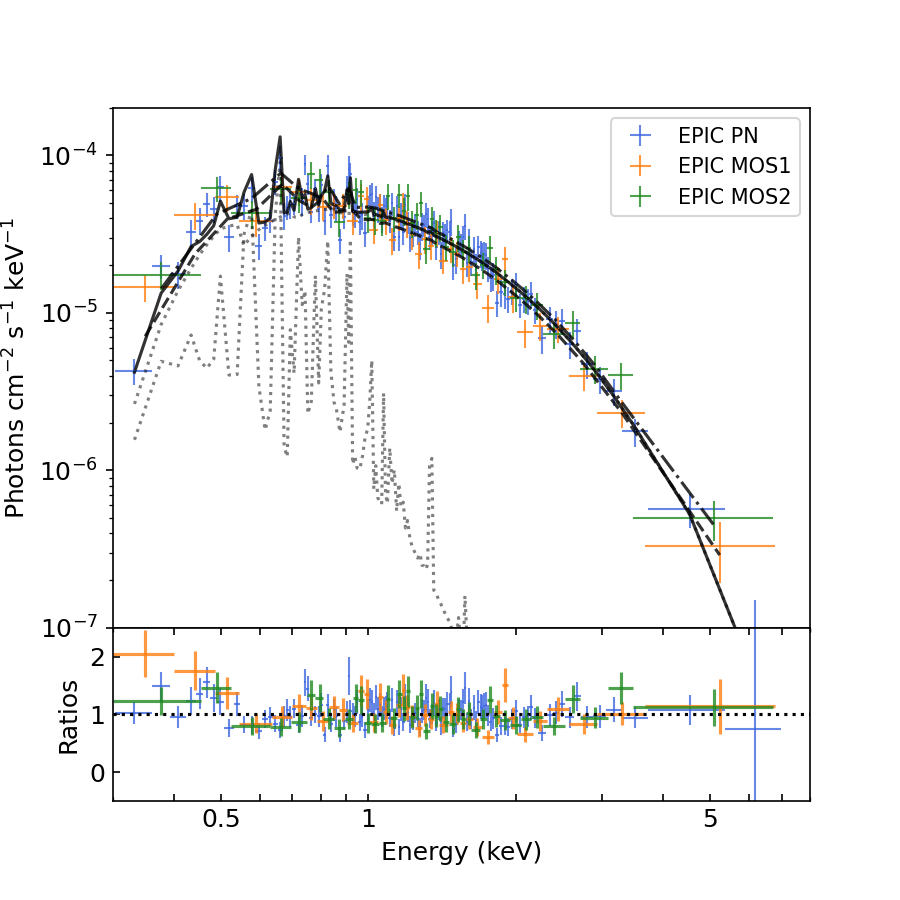}}\\    
    \caption{The favored spectroscopic models. (Top) Best-fitting spectral model and favored physical explanation of the observed X-ray emission, \textsc{tbabs$\times$(diskbb+pl)}. (Bottom) The \textsc{tbabs$\times$(diskbb+apec)} model, viewed as a potential alternative explanation for the observed emission. See  Section~\ref{sec:spectralfitting} for spectroscopic fitting process and the details on the best-fit parameters of these models.}
    \label{fig:spectrafits}
\end{figure}

The component normalizations in all models were left free to vary without constraints and the redshift was assumed to be $z=0.001797$ (when applicable). As with the three initial models described previously, we initially allowed the column density to be free to vary but froze it to the known Galactic value when the best fit models returned column densities below the known value. Thus, any column densities quoted in this work that exceed the known Galactic value should be considered the total line-of-sight column density (Galactic + extragalactic). In each of these cases, the added model component introduced two free parameters to the fit; as is commonly adopted, for each additional free parameter added to the model, we required the C-stat value of a fit to satisfy the criterion $\rm{Cstat}_{\rm{old}}-\rm{Cstat}_{\rm{new}}>2.71$ \citep[e.g.,][]{tozzi2006,brightman2014} in order to be considered a statistical improvement over the base \textsc{diskbb} model. In addition to this, we visually inspected the fit for each new model to ensure a proper fit was determined, and we used the \textsc{steppar} command in \textsc{xspec} to derive contour plots for parameters of interest; these plots were inspected to determine how well constrained the parameter values were for each fit. All three additional models offered statistically significant improvements over the simple multicolor disk model based on the change in the fit statistics (\textsc{diskbb+apec}: $\Delta$C-stat=59.23, \textsc{diskbb+pl}: $\Delta$C-stat=104.59, \textsc{diskbb+bbody}: $\Delta$C-stat=94.6, and a visual inspection of the fits suggested each could reasonably reproduce the observed spectrum.  

In the case of the multicolor disk + blackbody model (\textsc{diskbb+bbody}), the fit would suggest\footnote{It should be noted that an essentially identical fit (C-stat/dof = 1736.47) can be obtained where the situation is reversed if the initial parameters are both set to be 1\,keV and free to vary in the range 0.01-5\,keV: the model returns a hotter inner disk temperature of $T_{in}=0.66_{-0.03}^{+0.03}$\,keV and a cooler blackbody component at $0.13_{-0.02}^{+0.02}$\,keV. In this case the inner disk radius is found to be $R_{in}\,$cos$^{1/2}(i)$ = $101.5$\,km and the implied blackbody radius $R_{bb}=1589.5$\,km is still far far too large relative to that expected for emission from the surface a neutron star and is caused by the relatively small blackbody temperature in concert with the large normalization when calculating $R_{bb}$. This result is inconsistent with the hard blackbody component and soft multicolor disk component typical of neutron star X-ray binaries. Thus, this version of the model can also be ruled out.} a cooler inner disk temperature of $T_{in}=0.29_{-0.04}^{+0.04}$\,keV and a hotter blackbody component with $0.54_{-0.04}^{+0.05}$\,keV.  Although the bbody+diskbb model is commonly observed in bright LMXBs, the obtained temperatures are significantly lower than typical values of $kT_{bb}$ and $kT_{in}$ \citep[e.g.][]{mitsuda1984}, resulting in an unusually large blackbody radius of $R_{bb}=121.4^{+28.5}_{-28.2}$ km compared to that expected for a neutron star radius \citep[$\sim$13 km,][]{riley2019,miller2019}. While the \textsc{diskbb+bbody} model reproduces the observed continuum, we disfavor this model on physical grounds. 

\begin{table*}
\begin{center}
\caption{Spectral Fitting Results}
\label{apptable:specanalysis}
\begin{tabular}{ccccccccccccccc}
\hline
\hline
\noalign{\smallskip}
\noalign{\smallskip}
Model & C-stat & d.o.f & $N_{\rm{H}}$ & $\Gamma$ &  $\Gamma$ Norm. & T$_1$ \\
 &  &  & ($10^{22}$ cm$^{-2}$) &  & & (keV) & \\
(1) & (2) & (3) & (4) & (5) & (6) & (7) \\
\noalign{\smallskip}
\noalign{\smallskip}
\hline
\noalign{\smallskip}

$\rm{tbabs}\times \rm{diskbb}$ & 1831.07 & 1838 & $0.22^{*}$ & \dots & \dots & $0.56^{+0.02}_{-0.01}$  \\

$\rm{tbabs}\times \rm{pl}$ & 1815.72 & 1837 & $0.42^{+0.03}_{-0.03}$ & $2.97^{+0.09}_{-0.08}$ & $1.04^{+0.07}_{-0.07}\times10^{-4}$ & \dots \\ 

$\rm{tbabs}\times \rm{apec}^{\dagger}$ & 3370.81 & 1839 & $1.38_{-0.04}^{+0.03}$ & \dots & \dots & $0.69_{-0.03}^{+0.02}$ \\

$\rm{tbabs}\times (\rm{diskbb+bbody})^{\dagger\dagger}$ & 1736.18 & 1836 & 0.22$^{*}$ & \dots & \dots & $0.29^{+0.04}_{-0.04}$ \\ 


$\rm{tbabs}\times (\rm{diskbb+apec})$ & 1771.84 & 1836 & 0.22$^{*}$ & \dots & \dots & $0.61^{+0.02}_{-0.02}$ &   \\ 

$\rm{tbabs}\times (\rm{diskbb+pl})$ & 1726.48 & 1835 & $0.32^{+0.14}_{-0.13}$ & $3.48^{+1.08}_{-0.86}$ & $3.56^{+1.90}_{-2.09}\times10^{-5}
$ & $0.66^{+0.04}_{-0.06}$ \\ 


$\rm{tbabs}\times (\rm{thcomp \times diskbb})$ & 1728.57 & 1836 & $0.22^{*}$ & $1.9_{-0.03}^{+0.05}$ & \dots & $0.62_{-0.04}^{+0.04}$ \\

\noalign{\smallskip}
\hline

\noalign{\smallskip}
\noalign{\smallskip}
\noalign{\smallskip}
\noalign{\smallskip}
\hline
\hline
\noalign{\smallskip}
\noalign{\smallskip}
 T$_1$ Norm. & T$_2$ &  T$_2$ Norm. & $L_{\rm{2-10\,keV}}$ & $L_{\rm{0.3-10\,keV}}$ & $R_{in}$\,cos$^{1/2}\,i$ & $R_{bb}$\\
& (keV) & & (erg\,s$^{-1}$) & (erg\,s$^{-1}$) & (km) & (km) \\
(8) & (9) & (10) & (12) & (13) & (14) & (15)\\
\noalign{\smallskip}
\noalign{\smallskip}
\hline
\noalign{\smallskip}
$1.14^{+0.14}_{-0.12}\times10^{-1}$ & \dots & \dots & $6.23^{+0.14}_{-0.12}\times10^{37}$ & $2.11^{+0.03}_{-0.04}\times10^{38}$ & $144.9^{+8.5}_{-7.9}$ & \dots \\

\dots & \dots & \dots & $1.03^{+0.03}_{-0.03}\times10^{38}$ & $2.48^{+0.04}_{-0.04}\times10^{38}$ & \dots & \dots \\

$5.47^{+0.40}_{-0.35}\times10^{-4}$ & \dots & \dots & $4.21^{+0.07}_{-0.08}\times10^{37}$ & $1.78^{+0.02}_{-0.02}\times10^{38}$ & \dots & \dots \\

$1.08^{+0.73}_{-0.42}$ & $0.54^{+0.05}_{-0.04}$ & $1.25^{+0.18}_{-0.21}\times10^{-6}$ & $7.59^{+0.13}_{-0.27}\times10^{37}$ & $2.24^{+0.02}_{-0.07}\times10^{38}$ & $446.6^{+130.7}_{-97.7}$ & $121.4^{+28.5}_{-28.2}$ \\
%


$7.14^{+1.20}_{-1.07}\times10^{-2}
$ & $0.23^{+0.03}_{-0.03}$ & $2.22^{+0.53}_{-0.51}\times10^{-5}
$ & $7.12^{+0.15}_{-0.22}\times10^{37}$ & $2.2^{+0.02}_{-0.04}\times10^{38}$ & $114.5^{+9.3}_{-8.9}$ & \dots \\

$4.07^{+1.18}_{-1.17}\times10^{-2}$ & \dots & \dots & $7.82^{+0.35}_{-0.29}\times10^{37}$ & $2.27^{+0.05}_{-0.07}\times10^{38}$ & $86.4^{+17.3}_{-13.6}$ & \dots \\

\dots & $0.083_{-0.002}^{+0.071}$ & $110^{NC}_{-103}$ & $7.63^{+79.83}_{-4.40}\times10^{37}$ & $2.25^{+15.8}_{-2.30}\times10^{38}$ & $4495^{NC}_{-3387}$ & \dots \\

\noalign{\smallskip}
\hline
\end{tabular}
\end{center}
\tablecomments{Spectral fitting results from the various model choices. Col. 1: Model choice. Col. 2-3: C-stat and degrees of freedom of the model. Col. 4: total line-of-sight column density and photon index (when applicable). Col. 5-6: power law photon index ($\Gamma$) and component normalization (when applicable). Col. 7 and 9: temperatures of the \textsc{diskbb}, \textsc{bbody}, \textsc{apec}, and \textsc{thcomp} components (when applicable); note for the \textsc{tbabs}$\times$(\textsc{thcomp}$\times$\textsc{diskbb}) model, $T_1$ and $T_2$ here correspond to the \textsc{thcomp} electron temperature (kT$_e$) and \textsc{diskbb} inner disk temperature, respectively. Col. 8 and 10: normalizations of the respective thermal components given in Col. 7 and 9 (when applicable). Col. 12-13: observed $L_{2-10\,\rm{keV}}$ and $L_{0.3-10\,\rm{keV}}$ luminosities. Col 14-15: $R_{in}$\,cos$^{1/2}\,i$ and $R_{bb}$ values for the \textsc{diskbb} and \textsc{bbody} model components (when applicable). $\dagger$ indicates disfavored models based on C-Stat. $\dagger\dagger$ indicates disfavored models based on physical values. $*$ indicates the column density was frozen to the known Galactic value ($2.2\times10^{21}$ cm$^{-2}$). NC stands for ``Not Constrained'' and indicates a failure to constrain the error of a model component.
}
\end{table*}

 The best fit \textsc{diskbb+pl} model well reproduced the observed spectrum and returned an inner disk temperature of $T_{in}=0.66_{-0.06}^{+0.04}$\,keV and a rather steep photon index of $\Gamma=3.48_{-0.86}^{+1.08}$; the calculated inner disk radius $R_{in}\,$cos$^{1/2}(i)$ = $86.4_{-13.6}^{+17.3}$\,km is too large for a typical neutron star but is consistent with expectations for a black hole XRB. However, we note that a neutron star cannot be fully ruled out, since the magnetic field could truncate the accretion disk at large radii \citep[e.g.,][and references therein]{king2016}. Given the reasonable best-fit values, best-fit C-stat value, and the physically meaningful interpretation, we favor the \textsc{diskbb+pl} model as the physical interpretation of the source. Finally the multicolor disk and thermal plasma model (\textsc{diskbb+apec}) well reproduced the observed spectrum across the $\sim0.5-7$\,keV range, with an inner accretion disk temperature of $T_{in}=0.61_{-0.02}^{+0.02}$\,keV and a thermal plasma temperature of $kT=0.23_{-0.03}^{+0.03}$\,keV. Like in the case of the \textsc{diskbb+pl} model, the calculated inner disk radius $R_{in}\,$cos$^{1/2}(i)$ = $114.5_{-8.9}^{+9.3}$\,km is consistent with that expected for a black hole XRB and is too large for a neutron star XRB (except in the case of a truncated accretion disk). In this model case, the soft X-ray emission modeled by the \textsc{apec} component likely involves a complex combination of emission arising from the nearby halo of NGC 4945, the cosmic Galactic X-ray background, and instrumental background emission. In light of the reasonable inner disk radius and the potentially physically meaningful \textsc{apec} component, we view this as a viable alternative to the \textsc{diskbb+pl} model (though there is still some excess soft X-ray emission not accounted for by the \textsc{diskbb+apec} model). 
 
Motivated by the steep photon index derived in using the \textsc{diskbb+pl} model above, we replaced the power law component with the physical Comptonization model \textsc{thcomp} and iteratively refit the spectrum, freezing or thawing components based on the changes in reduced C-stat as described above for the other models.\footnote{Note: an initial fit with $N_{\rm{H}}$ frozen to the known Galactic value, covering factor frozen to 1, and $\rm{kT}_e$ frozen at 10 returned a nearly identical photon index ($\sim3.47$) to that obtained for the \textsc{diskbb+pl} model but a lower inner disk temperature of $\approx0.38$\,keV, while the derived the C-Stat value (1762) indicated a significantly poorer fit.} We converged on a model where the covering factor was frozen at 1, $N_{\rm{H}}$ was frozen to the known Galactic value ($2.2\times10^{21}$\,cm$^{-2}$), while the power law index, electron temperature, black body temperature, and normalization were left free to vary. This model (\textsc{tbabs$\times$(thcomp$\times$diskbb)}) returned a flatter power law index ($\Gamma=1.9_{-0.03}^{+0.05}$), electron temperature $\rm{kT}_e$ = $0.62_{-0.04}^{+0.04}$ keV, inner accretion disk temperature $T_{in}=0.083 _{-0.002}^{+0.071}$\,keV, and resulted in a C-Stat value of 1728.6 with 1836 degrees of freedom.

This model proved rather unstable during the fitting process, however; the reported values here were derived during the calculation of error within \textsc{xspec}, during which the normalization of the \textsc{diskbb} component was always found to rapidly increase as the blackbody temperature decreased. As a result, while this model can reproduce the observed spectrum, the normalization cannot be precisely constrained. A visual inspection of the ``best fit'' for the \textsc{thcomp$\times$diskbb} model revealed that the \textsc{thcomp} component on its own could reproduce the shape of the continuum while the \textsc{diskbb} component contributed negligibly; for this model, XMM-Newton lacks sufficient soft X-ray coverage to properly constrain the blackbody properties. For comparison to the above estimates, this model suggests an inner accretion disk radius of $R_{in}\,$cos$^{1/2}(i)$ = $4495_{-3387}^{+NC}$\,km, far exceeding any other $R_{in}\,$cos$^{1/2}(i)$ value derived in this work.  Given the slightly higher C-Stat value and instability of this model, we still favor the \textsc{diskbb+pl} model. 

 Throughout the remainder of this work, we quote the \textsc{diskbb+pl} model as the best-fit and favored model, and we also occasionally quote the best-fit parameters for the \textsc{diskbb+apec} and \textsc{thcomp$\times$diskbb} model for comparison purposes. Table~\ref{apptable:specanalysis} provides the best fitting parameters for the models described in this section, and the favored spectral models are shown in Figure~\ref{fig:spectrafits}. Following the analysis of the $\textsc{thcomp}\times\textsc{diskbb}$, it is evident that the steep power law component found with the $\textsc{diskbb}\times\textsc{pl}$ model appears as an excess at lower energies rather than a high energy tail in the spectrum. This is reminiscent of the ultrasoft state of XTE J1550–564 \citep{kubota2004} or the soft state of ULXs \citep[e.g., M81 X-9;][]{tsunoda2006}, where the steep power law has been interpreted as compensating for deviations from the standard Shakura–Sunyaev disk \citep[][with T $\propto$ r$^{-0.75}$]{shakura1973}, and suggests that the accretion disk is in a slim disk-like state (T $\propto$ r$^{-0.5}$).

The observed 0.3-10\,keV and 2-10\,keV luminosities derived from the best-fitting models described above are consistent with that expected for an XRB but low relative to what might be expected for a ULX: the \textsc{diskbb+pl} model yielded $L_{0.3-10\,\rm{keV}}=2.27^{+0.05}_{-0.07}\times10^{38}$ erg s$^{-1}$ and $L_{2-10\,\rm{keV}}=7.82_{-0.29}^{+0.35}\times10^{37}$ erg s$^{-1}$, while the \textsc{diskbb+apec} model yielded $L_{0.3-10\,\rm{keV}}=2.2^{+0.02}_{-0.04}\times10^{38}$ erg s$^{-1}$ and $L_{2-10\,\rm{keV}}=7.12_{-0.22}^{+0.15}\times10^{37}$ erg s$^{-1}$.  If we assume that the inner disk radius represents the innermost stable circular orbit (ISCO) and make the simple assumption that the object is non-spinning, we can derive the central mass of the object via $R_{in}=3R_S=6\rm{GM}/\rm{c}^2$. With $R_{in}=86.4_{-13.6}^{+17.3}\,$cos$^{-1/2}(i)$\,\,km from the \textsc{diskbb+pl} model, we find a central mass of $9.8^{+1.0}_{-0.8}$\,cos$^{-1/2}(i)$\,\,M$_\odot$, suggestive of a black hole rather than a neutron star; for an inclination angle of 60\deg, this in turn yields a (non-spinning) black hole mass of $13.9^{+1.4}_{-1.1}$\,M$_\odot$.

All models involving a \textsc{diskbb} component returned similar masses, in the range of $\sim(9.8-14.7)$\,cos$^{-1/2}(i)$\,M$_{\odot}$ (the \textsc{diskbb+apec} model yielding $12.9^{+0.5}_{-0.5}$\,cos$^{-1/2}(i)$\,M$_{\odot}$, or $18.2^{+0.7}_{-0.7}$\,M$_\odot$ assuming an inclination angle of 60\deg), with the exception of the \textsc{thcomp$\times$diskbb} model which returned a mass of $\sim500$\,cos$^{-1/2}(i)$\,M$_\odot$, likely due to the insensitivity of the \textsc{diskbb} component to the fit. Note: a neutron star cannot be ruled out at this point due to the fact that neutron star magnetic fields can truncate the accretion disk at large radii \citep[e.g.,][]{papitto2010,papitto2013,king2016}, and in such a case our assumption regarding the inner disk radius and ISCO would be incorrect and would lead to an overestimation of the compact accretor's mass. 

We can similarly use the inner accretion disk temperature  and the normalization of the multicolor disk component to compute the bolometric luminosity of the source: as outlined in \citet{mitsuda1984}, the apparent inner disk radius, $r_{in}$, is derived from the normalization of the multicolor disk component, and the bolometric luminosity of the accretion disk is thus given by the expression $L_{bol}=4\pi r_{\rm{in}}^2\sigma T_{\rm{in}}^4$, where $\sigma$ is the Stefan-Boltzmann constant. This expression yields a bolometric luminosity of $1.27^{+0.65}_{-0.56}\times10^{38}{\,\rm{cos}}^{-1/2}(i)$\,erg\,s$^{-2}$ when assuming the best fit parameters from the  \textsc{tbabs$\times$(diskbb+pl)} model.

\section{Insights from Swift and WISE}

\subsection{Searching for the Source Activation with Swift}
\label{sec:swift}
\begin{figure}
    \centering
    \includegraphics[width=\linewidth]{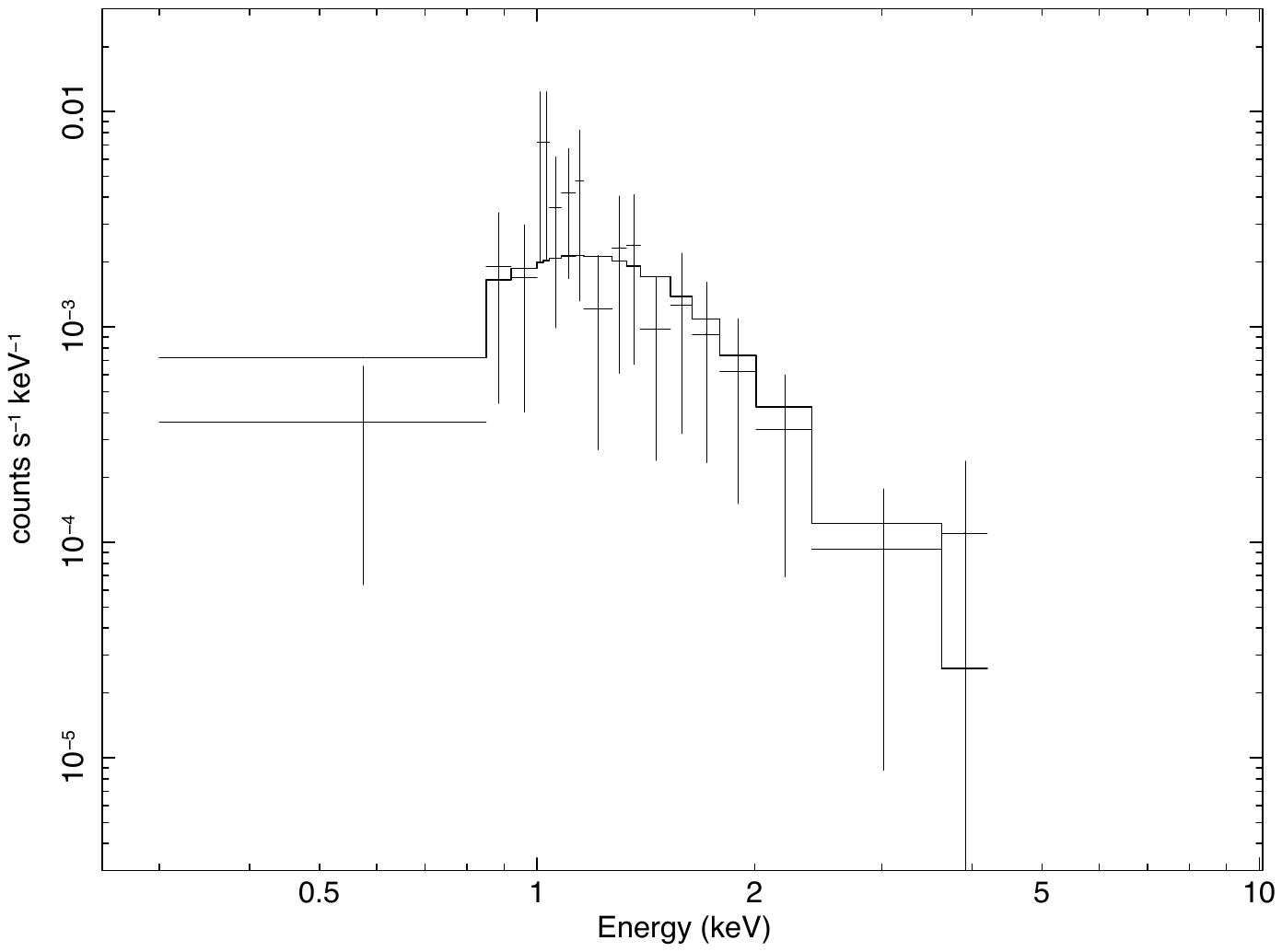}\\
    \includegraphics[width=\linewidth]{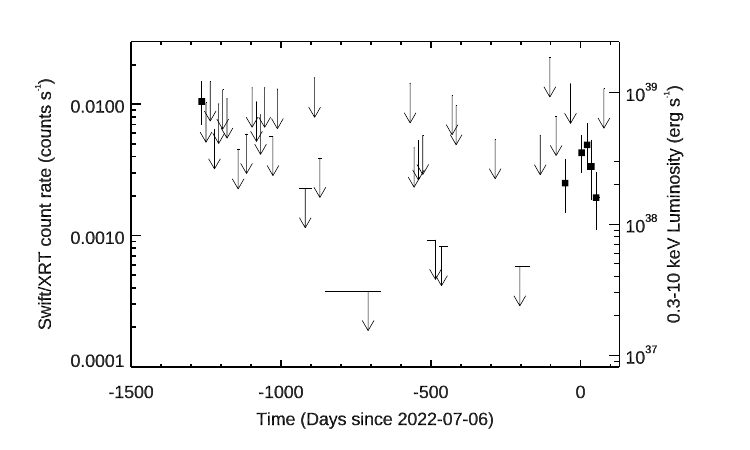}
    \caption{(Top) The stacked Swift-XRT 0.3-10\,keV X-ray spectrum of XMM J130514.64-493311.27, binned at 1 count per bin. Total exposure time: 13.7\,ks. The apparent lack of emission above $\sim4$\,keV is likely due to the softer nature of the source in combination with the relatively short exposure. The best-fit absorbed multi-color disk model is depicted with a solid black line (see Section~\ref{sec:swift} for further details). (Bottom) Swift-XRT light curve of XMM J130514.64-493311.27 in NGC 4945. Upper limits (3$\sigma$) are shown with arrows. The luminosity axis on the right assumes a distance of 3.7 Mpc to the source and the count rate to flux conversion factor described in Section~\ref{sec:swift}. Note: the 2008 Swift-XRT detection falls outside of the figure axes.}
    \label{fig:swift_ltcrv}
\end{figure}

The Neil Gehrels Swift Observatory \citep{2004ApJ...611.1005G} has observed NGC 4945 146 times between 2008-03-07 and 2022-10-02 (Target IDs 13908, 14065, 15017, 32227, 37266, 49859, 80104, 84458, 00572489 and 03109090). We used the online tool provided by the University of Leicester\footnote{https://www.swift.ac.uk/user\_objects/} \citep{2007A&A...469..379E,2009MNRAS.397.1177E} to generate the Swift light curve and spectrum. We did not allow centroiding of the data since this is known to fail for faint sources such as this. XMM J130514.64-493311.27 was only detected in 7 of the Swift/XRT observations, once on 2008-04-29, once on 2019-01-19, and 5 times between 2022-05-15 and 2022-08-30. Our XMM-Newton imaging coincides with this latest period of source activity. We show the light curve along with upper limits in Figure \ref{fig:swift_ltcrv}. 

Using the online tool provided by the University of Leicester, we extracted the stacked 0.3-10\,keV Swift-XRT spectrum of XMM J130514.64-493311.27 -- shown in Figure~\ref{fig:swift_ltcrv} -- which had a total exposure time was 13.7 ks. We fit the spectrum with two simple models: an absorbed power law model ($\textsc{tbabs}\times\textsc{pl}$) and an absorbed multicolor disk model ($\textsc{tbabs}\times\textsc{diskbb}$), where the $N_{\rm{H}}$ was frozen to the known Galactic value \citep[$N_{\rm{H},\,\rm{Gal.}}=2.2\times10^{21}$\,cm$^{-2}$;][]{willingale13} and the models were kept simple due to the low number of counts ($<50$) relative to the XMM spectrum. The power law model returned a $\rm{C}_{\rm{stat}}=30.93$ with 35 DoFs, $\Gamma=2.35^{+0.57}_{-0.60}$, and a 0.3-10 keV absorbed flux of $1.21^{+0.48}_{-0.23}\times10^{-13}$\,erg\,cm$^{-2}$\,s$^{-1}$. At 3.6 Mpc, this flux implies a 0.3-10 keV luminosity of $1.88^{+0.74}_{-0.36}\times10^{38}$\,erg\,s$^{-1}$; the photon index and luminosity values returned by this model are consistent with the XMM spectrum within the errors. The multi-color disk model returned $C_{\rm{stat}}=27.84$ with 35 DoFs, a thermal temperature $T=0.52^{+0.24}_{-0.14}$\,keV (consistent with the multi-color disk fit to the XMM data), and a 0.3-10 keV absorbed flux of $8.08^{+0.50}_{-5.05}\times10^{-14}$\,erg\,cm$^{-2}$\,s$^{-1}$. At 3.6 Mpc, the source exhibits a 0.3-10 keV luminosity of $1.25^{+0.08}_{-0.78}\times10^{38}$\,erg\,s$^{-1}$; this implies an unabsorbed 0.3-10 keV luminosity of $2.20\times10^{38}$\,erg\,s$^{-1}$. The count rate to flux conversion factor was $4.98\times10^{-11}$ erg cm$^{-2}$ count$^{-1}$, which we used to determine the luminosity axis in Figure \ref{fig:swift_ltcrv}.



\subsection{Searching for A Multiwavelength Counterpart to XMM J130514.64-493311.27}
\label{sec:multiwavelengthsource}

We also searched for multiwavelength counterparts to XMM J130514.64-493311.27. Using the Wide-field Infrared Survey Explorer All-Sky Survey, we searched for variability in the mid-infrared using a custom Python code \citep[see][for a description]{secrest2020}. In short, this code compiles W1 and W2 magnitudes for a given target in the ALLWISE and NEOWISE catalogs, and tests for source variability through several non-parametric metrics, including those described in \citet{kozlowski2016}, \citet{sesar2007}, and \citet{rakshit2019}. NEOWISE detected a potential mid-infrared counterpart within 3'' of the X-ray position (with S/N$\geq3$ in W1) 12 times between January 2015 and 2023; the earliest detection exhibited a W1 magnitude of 15.91$\pm$0.22, and the source reached a peak brightness of W1mag = 15.46$\pm$0.20 in July 2018. XMM J130514.64-493311.27 was consistently detected in W1 at least once per year in 2015-2018, 2020-2022, and in 2023, but the source was not detected in the W2 band. XMM J130514.64-493311.27 does appear to be fading in the mid-IR below the observed mid-IR flux levels in 2015-2016. There is no apparent correlated activity between the mid-IR and X-ray emission from May-August 2022 (based on a single WISE observation) when the source was detected in the X-rays by XMM-Newton in July and 5 times by Swift-XRT between May and August. There unfortunately was no WISE observation at the time of the 2019 Swift-XRT X-ray detection, so it is unclear if there was any correlated activity at that time. Conversely, there were no X-ray detections by Swift-XRT during the periods of WISE activity. Due to the cadence of WISE imaging, it is unclear if the mid-IR detections between Jan. 2016 and July 2021 correspond to distinct outbursts of activity or some level of sustained activity.

In the context of black hole low mass X-ray binaries (LMXBs), outbursts are expected to take place on timescales of 50-100 days \citep[e.g.,][]{john2024}. Prior works have shown that such systems can show episodes of X-ray-faint brightening where mid-IR counterparts are detected but the source remains undetected in the X-rays \citep{john2024,alabarta2021} and that such episodes can be attributable to failed transition outbursts \citep{alabarta2021}. The X-ray-undetected outbursts in a sample of LMXBs discussed by \citet{john2024} were associated with flat/inverted broadband mid-IR spectra (where the W1 flux was approximately equal to or greater than the W2 flux); the lack of W2 detections for the mid-IR counterpart in this work potentially suggests a flat/inverted spectrum. The prior mid-IR emission and lack of X-ray detections (obtained non-simultaneously) could be interpreted as XMM J130514.64-493311.27 being a black hole LMXB and undergoing a series of FT outbursts across 2016-2021, but the lack of better temporal cadencing in the mid-IR and X-ray light curves makes it difficult to draw any firm conclusions.

Black hole LMXBs during a typical outburst tend to show correlated X-ray and mid-IR emission \citep{john2024}; XMM J130514.64-493311.27 does not appear to show a correlation between the X-rays and mid-IR (based on a single WISE observation) during the X-ray-bright outburst between May 2022-August 2022, however the poor mid-IR temporal cadencing makes it difficult to draw firm conclusions.

No counterpart sources were found in the ALLWISE or CatWISE2020 catalogs. We have found no counterpart at other wavelengths, spanning the near-IR (2MASS), UV (GALEX), optical (DSS, HST, Gaia), or radio (VLASS 2.1).

\section{Results and Discussion}
\label{sec:summ_discuss}

In this work, we have identified a new X-ray transient, XMM J130514.64-493311.27, in the deepest XMM-Newton X-ray observation of the NGC 4945 obtained to-date (obtained July 2022). Located at $\alpha=13\hr{}05'14.\hspace{-1mm}''64$, $\delta=-49\deg{}33'11.\hspace{-1mm}''27$, the source lies $\sim5.5$' southeast of the galaxy nucleus, residing either along or below the plane of the galaxy. The transient is relatively soft (HR=-0.73) and shows no evidence for flaring across the $\sim99$\,ks pn exposure. We found no evidence for XMM J130514.64-493311.27 in the 2001 or 2004 XMM-Newton imaging, nor in any previous Chandra observations, including a mosaic stack of Chandra observations spanning 2000-2021. We could not identify any multiwavelength counterpart to this new source in the AllWISE, CatWISE2020, 2MASS, and GALEX point source catalogs nor in DSS and VLASS imaging, but we did identify a mid-IR counterpart in NEOWISE mid-IR light curves.  Based on archival Swift-XRT observations of NGC 4945, XMM J130514.64-493311.27 appears to have been active at these X-ray flux levels during only three short periods since early 2008 (one detection in 2008, one detection in 2019, five detections between May and August 2022); our new XMM-Newton imaging traces the most recent period of X-ray activity. The NEOWISE mid-IR counterpart was detected in several WISE observations spanning 2015-2018, 2020-2022, and once in 2023; none of these mid-IR detections overlap with the X-ray active periods.

The X-ray spectrum of XMM J130514.64-493311.27 can be well described by a multicolor disk + power law model (\textsc{tbabs$\times$[diskbb+pl]}, with $\Gamma=3.48_{-0.86}^{+1.08}$ and inner disk temperature $0.66_{-0.06}^{+0.04}$\,keV) as well as a multicolor disk + thermal plasma model (\textsc{tbabs$\times$[diskbb+apec]}, with inner disk temperature $0.61_{-0.02}^{+0.02}$\,keV and thermal plasma temperature $0.23_{-0.03}^{+0.03}$\,keV). These models yield inner disk radii of $R_{in}\,$cos$^{1/2}(i) = 86.4_{-13.6}^{+17.3}$\,km and $R_{in}\,$cos$^{1/2}(i) = 114.5_{-8.9}^{+9.3}$\,km, respectively, and 0.3-10\,keV luminosities in the range of $\sim2.2-2.3\times10^{38}$ erg $^{-1}$. We favor the \textsc{tbabs$\times$(diskbb+pl)} model as the physical interpretation of the observed X-ray emission, where the multicolor disk models the disk emission of the accretor while the power law component models Compton up-scattered emission from a corona. We view the \textsc{diskbb}+\textsc{apec} model as an alternative interpretation of the source, where the  multicolor disk again models the accretion disk while the thermal plasma models the soft X-ray emission due to a combination of the nearby galaxy halo emission and soft X-ray background emission. The best-fit inner accretion disk temperature and derived inner disk radius for the \textsc{tbabs$\times$(diskbb+pl))} model yields a source bolometric luminosity of $1.27^{+0.65}_{-0.56}\times10^{38}{\,\rm{cos}}^{-1/2}(i)$\,erg\,s$^{-1}$. If we relate the inner accretion disk radius to the innermost stable orbit of a non-spinning compact object, we obtain a central mass of $9.8^{+1.0}_{-0.8}$\,cos$^{-1/2}(i)$\,\,M$_{\odot}$, implying the presence of a stellar mass black hole, but a neutron star with a magnetically truncated disk cannot be ruled out.




Our detection of a few periods of X-ray activity in archival Swift-XRT observations dating back to 2008 is consistent with XMM J130514.64-493311.27 undergoing distinct outbursts and the lack of detections in archival Chandra and XMM-Newton imaging. This long-term variability is also consistent with the lack of multiwavelength counterparts in most archival point source catalogs (e.g., AllWISE, 2MASS, etc.); the mid-IR counterpart identified in NEOWISE light curves may suggest additional outbursts or failed outbursts \cite[e.g.,][]{alabarta2021}. The lack of an obvious multiwavelength counterpart in, for example, optical imaging argues against a background object such as an AGN in a more distant galaxy. The off-nuclear position of XMM J130514.64-493311.27, lack of a persistent or correlated multiwavelength counterpart, and long-term X-ray variability argues in favor of an ultraluminous X-ray source (ULX) or an X-ray binary (XRB) for the nature of XMM J130514.64-493311.27. Typical ULXs show peak luminosities on the order of a few $10^{39}$ up to $10^{41}$ erg s$^{-1}$ \citep[e.g.,][]{makishima2000}. XMM J130514.64-493311.27 instead exhibits an observed 0.3-10\,keV X-ray luminosity of $L_{0.3-10\,\rm{keV}}=2.27^{+0.05}_{-0.07}\times10^{38}$ erg s$^{-1}$, based on the \textsc{tbabs$\times$(diskbb+pl)} model fit to our XMM-Newton observation, roughly an order of magnitude lower than that expected for a ULX. The Swift-XRT stacked 13.7\,ks spectrum on its own would suggest that XMM J130514.64-493311.27 could be an XRB or a ULX depending on the model ($L_{0.3-10\,\rm{keV}}\sim0.3-6\times10^{39}$ erg s$^{-1}$), but this could be explained by the Swift-XRT spectrum having fewer spectral counts; the total Swift-XRT exposure is $\sim7\times$ shorter than the new XMM-Newton observation. 

All of this evidence taken together - the long-term X-ray variability, off-nuclear position, lack of persistent/correlated multiwavelength counterpart, low X-ray luminosity, as well as the softness of the source and inner disk temperatures and radii derived from the spectra, suggests that XMM J130514.64-493311.27 is a newly-detected XRB in NGC 4945 rather than a new ULX. \textcolor{black}{The derived inner disk radius ($R_{in}\,$cos$^{1/2}(i) = 86.4_{-13.6}^{+17.3}$\,km ) and the implied mass of the compact accretor ($\approx9.8$\,cos$^{-1/2}(i)$\,\,M$_{\odot}$) suggests that XMM J130514.64-493311.27} is powered by a black hole rather than a neutron star, although we still cannot rule out that the compact object is a neutron star. A neutron star's magnetic field can truncate the accretion disk at large radii \citep[e.g.,][]{papitto2010,papitto2013,king2016}, and in such a case the accretion disk would not extend to the ISCO like we assumed in our mass estimation in Section~\ref{sec:spectralfitting}. Interestingly, XMM J130514.64-493311.27 is significantly displaced from the galaxy, and appears to reside (in projection) in the direction of the recently detected iron K$\alpha$ nebula extending $\sim220$'' in a direction southeast from the nucleus \citep{weaver2024}, as well as an extension of the H~I halo to the southeast of the galaxy disk \citep{Ianjamasimanana2022}. Further examination of this new transient would benefit from deep optical imaging to search for an optical counterpart, and additional X-ray imaging to monitor the source flux over time. 

\acknowledgments
We thank the anonymous referee for their careful review of this work, which helped to improve the quality and content of this manuscript. R. W. P. and J. M. C. gratefully acknowledge support through an appointment to the NASA Postdoctoral Program at Goddard Space Flight Center, administered by ORAU through a contract with NASA. M. M. gratefully acknowledges NASA’s support under award number 80GSFC24M0006. We thank H. Earnshaw, T. Piro, and E. Nathan for insightful discussions.

The scientific results reported in this article are based in part on data obtained from obtained by the Chandra X-ray Observatory, contained in~\dataset[DOI: 10.25574/cdc.552]{https://doi.org/10.25574/cdc.552}. This research has made use of software provided by the \textit{Chandra} X-ray Center (CXC) in the application packages \textsc{CIAO}. This research has made use of data obtained from the Chandra Source Catalog, provided by the Chandra X-ray Center (CXC). This research has made use of the NASA/IPAC Infrared Science Archive, which is funded by the National Aeronautics and Space Administration and operated by the California Institute of Technology. This publication makes use of data products from the Wide-field Infrared Survey Explorer, which is a joint project of the University of California, Los Angeles, and the Jet Propulsion Laboratory/California Institute of Technology, funded by the National Aeronautics and Space Administration.
\vspace{12mm}
\facilities{\chandra{}, \xmm{}, \textit{Swift}-XRT, \wise{}, Sloan, CTIO:2MASS, FLWO:2MASS, IRAS}

\software{APLpy \citep{robitaille2012}, pandas \citep{mckinney2010}, NumPy \citep{oliphant2006,walt2011,harris2020array}, SciPy \citep{virtanen2020}, \textsc{matplotlib} \citep{hunter2007}, \textsc{heasoft} \citep{heasoft}, \textsc{xspec} \citep{arnaud1996}, \textsc{ciao} \citep{fruscione2006}, \textsc{sas} \citep{2004ASPC..314..759G}, \textsc{ds9} \citep{joyce2003}, \textsc{Astropy} \citep{2013A&A...558A..33A,2018AJ....156..123A}}\\



\bibliography{references}{}
\bibliographystyle{aasjournal}

\end{document}